\documentclass[letterpaper, 10 pt, conference]{ieeeconf}  

\IEEEoverridecommandlockouts                              

\IEEEoverridecommandlockouts                              

\usepackage{graphics} 
\usepackage{amsmath} 
\usepackage{amssymb}  

\newcommand{\ba}{\begin{array}}
	\newcommand{\ea}{\end{array}}
\newcommand{\babc}{\begin{abc}}
	\newcommand{\eabc}{\end{abc}}
\newcommand{\bc}{\begin{center}}
	\newcommand{\ec}{\end{center}}
\newcommand{\be}{\begin{equation}}
\newcommand{\ee}{\end{equation}}
\newcommand{\bea}{\begin{eqnarray}}
\newcommand{\eea}{\end{eqnarray}}
\newcommand{\beas}{\begin{eqnarray*}}
	\newcommand{\eeas}{\end{eqnarray*}}
\newcommand{\bh}{\begin{hangitem}}
	\newcommand{\eh}{\end{hangitem}}
\newcommand{\bhi}{\begin{hangitem}}
	\newcommand{\ehi}{\end{hangitem}}
\newcommand{\bi}{\begin{itemize}}
	\newcommand{\ei}{\end{itemize}}
\newcommand{\bn}{\begin{enumerate}}
	\newcommand{\en}{\end{enumerate}}
\newcommand{\bq}{\begin{quote}}
	\newcommand{\eq}{\end{quote}}
\newcommand{\btb}{\begin{tabular}}
	\newcommand{\etb}{\end{tabular}}
\def\litem[#1]{\item[#1\hfill]}         
\newcommand{\abs}[1]{\left\vert #1 \right\vert}

\usepackage{csquotes}
\usepackage{multirow}
\usepackage{cite}
\usepackage{float}
\usepackage{array}
\newcolumntype{L}[1]{>{\raggedright\let\newline\\\arraybackslash\hspace{0pt}}m{#1}}
\newcolumntype{C}[1]{>{\centering\let\newline\\\arraybackslash\hspace{0pt}}m{#1}}
\newcolumntype{R}[1]{>{\raggedleft\let\newline\\\arraybackslash\hspace{0pt}}m{#1}}
\usepackage{color}

\def\QEDclosed{\mbox{\rule[0pt]{1.3ex}{1.3ex}}} 

\def\QED{\QEDclosed} 

\usepackage{graphicx}
\usepackage{subfig}
\usepackage{float}
\usepackage{amsmath}
\usepackage{subfiles}
\usepackage{enumerate}
\usepackage[T1]{fontenc}
\usepackage{amssymb}
\usepackage{mathtools}
\usepackage{color}
\usepackage{soul}

\usepackage{algorithm}
\usepackage{algorithmicx}
\usepackage{algpseudocode}

\usepackage{ntheorem}

\newtheorem{remark}{Remark}
\newtheorem{assumption}{Assumption}
\newtheorem{definition}{Definition}

\algdef{SE}[DOWHILE]{Do}{doWhile}{\algorithmicdo}[1]{\algorithmicwhile\ #1}

\renewcommand{\baselinestretch}{0.9845}

\title{\LARGE \bf Multi-Frequency Canonical Correlation Analysis (MFCCA): A Generalised Decoding Algorithm for Multi-Frequency SSVEP}

\author{Jing Mu, Ying Tan, David B. Grayden, and Denny Oetomo%
\thanks{This work was supported by the Valma Angliss Trust.}
\thanks{J. Mu, Y. Tan, and D. Oetomo are with the Department of Mechanical Engineering, The University of Melbourne, Parkville, VIC 3010, Australia.}
\thanks{D. B. Grayden is with the Department of Biomedical Engineering, The University of Melbourne, Parkville, VIC 3010, Australia.}
\thanks{{\tt\small j.mu@student.unimelb.edu.au, \{grayden, yingt, doetomo\}@unimelb.edu.au}}
}

\begin{document}

\maketitle
\thispagestyle{empty}
\pagestyle{empty}

\begin{abstract}
Stimulation methods that utilise more than one stimulation frequency have been developed for steady-state visual evoked potential (SSVEP) brain-computer interfaces (BCIs) with the purpose of increasing the number of targets that can be presented simultaneously. However, there is no unified decoding algorithm that can be used without training for each individual users or cases, and applied to a large class of multi-frequency stimulated SSVEP settings. This paper extends the widely used canonical correlation analysis (CCA) decoder to explicitly accommodate multi-frequency SSVEP by exploiting the interactions between the multiple stimulation frequencies.
A concept of order, defined as the sum of absolute value of the coefficients in the linear combination of the input frequencies, was introduced to assist the design of Multi-Frequency CCA (MFCCA).
The probability distribution of the order in the resulting SSVEP response was then used to improve decoding accuracy.
Results show that, compared to the standard CCA formulation, the proposed MFCCA has a 20\% improvement in decoding accuracy on average at order 2, while keeping its generality and training-free characteristics.

\end{abstract}


\section{Introduction}

Steady-state visual evoked potential (SSVEP) is a widely used signal for electroencephalography (EEG)-base brain-computer interfaces (BCIs) for its high information transfer rate (ITR) and minimum training requirements \cite{nicolas2012brain}.
In the past decade, multi-frequency stimulation has been explored to increase its potential of presenting larger numbers of targets needed. However, even though there are already several multi-frequency stimulation methods available, the decoding algorithms are still employed on an ad-hoc basis.

The existing decoding methods used in multi-frequency stimulated SSVEP are either frequency domain magnitude analysis, or based on prior knowledge of the resulting SSVEP, in other words, training-based approach.
Power spectral density based analysis (looking at magnitudes in frequency domain) is used in \cite{shyu2010dual,hwang2013new}. In \cite{chang2014amplitude}, the decoding algorithm is modified for each subject, hence requires training data. The same stimulation method is used in \cite{chen2013brain,chen2017novel}, however, the decoding algorithm is formulated differently in the two papers. This difference may come from the different set of frequencies selected in the two experiments, therefore requires prior knowledge to the resulting SSVEP. A training-based decoding algorithm is used in \cite{liang2020optimizing}.
A systematic approach that suits a range of stimulation setups and does not require individual training may improve the performance of the BCI.

Here, we introduce a generalised training-free approach to identify multi-frequency SSVEP. With this approach, a widely used decoding algorithm in SSVEP -- Canonical Correlation Analysis (CCA) was extended to Multi-Frequency CCA (MFCCA) based on two assumptions and the concept of ``order'' in multi-frequency stimulated SSVEPs. 
This algorithm could be applied to every multi-frequency stimulated SSVEP that meets these two assumptions.
In this paper, the assumptions are validated and the effectiveness of this algorithm is demonstrated using experimental results.

\section{Methods}

\subsection{Multi-Frequency Canonical Correlation Analysis (MFCCA) Formulation}

CCA compares correlations between the measured multi-channel EEG data $x(t)$ and a reference set of sinusoidal signals at the fundamental frequencies and harmonics defined as \cite{lin2006frequency}
\begin{equation}\label{eq:y_single}
    y(t)=
    \begin{bmatrix}
        \sin(2\pi ft)\\
        \cos(2\pi ft)\\
        \vdots \\
        \sin(2\pi N_h ft)\\
        \cos(2\pi N_h ft)
    \end{bmatrix},
\end{equation}
where $f$ is the fundamental frequency, $N_h$ is the number of harmonics in the reference, and $t$ is time.
The candidate frequencies are used to formulate the reference sets. The candidate that results in the highest correlation with $x(t)$ is the decoded output. 

\subsubsection{Direct Extension of CCA to Multi-Frequency}
The standard CCA works with stimulation that has only one input frequency. In the case where there are multiple frequencies as input, this standard CCA could be directly extended to a multi-frequency case by cascading the reference sets of each input frequency. Therefore, the multi-frequency reference set becomes
\begin{equation}\label{eq:y_multi}
    y_{\textrm{multi}}(t)=
    \begin{bmatrix}
        y_1(t)\\
        y_2(t)\\
        \vdots \\
        y_N(t)
    \end{bmatrix},\ 
    y_i(t)=
    \begin{bmatrix}
        \sin(2\pi f_i t)\\
        \cos(2\pi f_i t)\\
        \vdots \\
        \sin(2\pi N_h f_i t)\\
        \cos(2\pi N_h f_i t)
    \end{bmatrix},
\end{equation}
where $N_h$ is the number of harmonics, $f_i$, $i=1,2,\dots,N$ is the $i^{th}$ stimulation frequency, and $N$ is the number of frequencies.

This formulation could work in the case where the input frequencies are relatively independent and clear peaks can be observed at the input frequencies and their harmonics. However, from the literature, some interactions between the input frequencies are also clearly observed in the resulting SSVEP \cite{shyu2010dual,chang2014amplitude,chen2013brain,mu2021frequency}; in \cite{hwang2013new}, only the sum of the input frequencies was clearly observed in the SSVEP. Therefore, understanding the most common feature in multi-frequency SSVEP can help with the design of a generalised decoding algorithm.

\begin{figure}
    \centering
    \vspace{0.2cm}
    \includegraphics[width=0.65\linewidth, keepaspectratio]{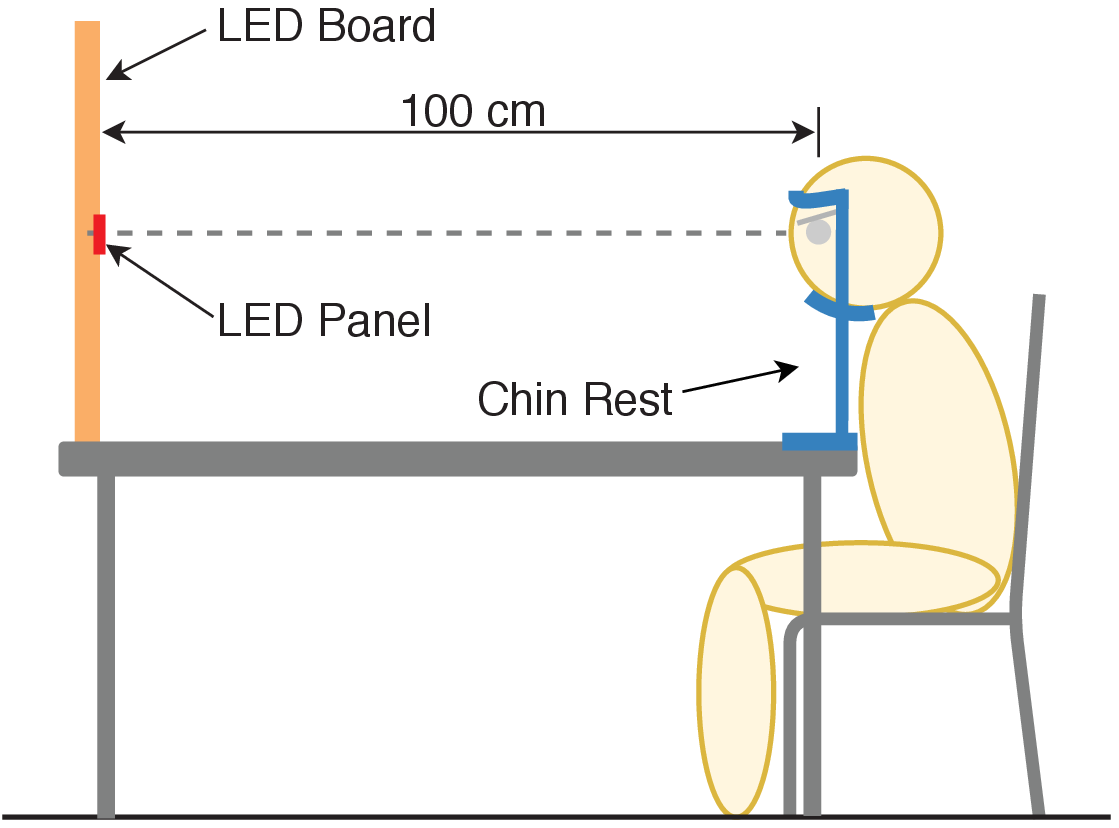}
    \caption{Experimental setup. A chin rest helps maintaining a 100 cm distance between the subject and the LED board, and keeping the centre of the LED board at eye level in the sagittal plane of the subject.}\vspace{-0.3cm}
    \label{fig:Experiment}
\end{figure}

\subsubsection{Assumptions on Common Multi-Frequency SSVEP Features}\label{sec:assumptions}

Based on the observations from our preliminary study and the literature \cite{shyu2010dual,hwang2013new,chang2014amplitude,chen2013brain,mu2021frequency}, two assumptions and one definition are introduced. The Multi-Frequency CCA (MFCCA) method was developed based on these assumptions and definition.

\begin{assumption}
The frequency components in the resulting SSVEP response contain the stimulation frequencies and their harmonics as well as frequencies that are linear combinations of the input frequencies with integer coefficients.
\end{assumption}

The concept of the ``order'' is defined as the sum of absolute values of the coefficients in a linear combination, essentially indicating the number of terms used in the reference set:
\begin{definition}
Assume that a signal $z(t)$ can be represented as a linear combination of basis signals $z_i(t)$; i.e., $z(t)=\sum_{i=1}^{m} a_i z_i(t)$, where $a_i\in \mathbb{Z}$ are coefficients; $\mathbb{Z}$ represents the set of all integers.
The `order' of $z(t)$ is defined as $N_\mathrm{O}=\sum_{i=1}^{m}\abs{a_i}$, where $N_\mathrm{O}\in \mathbb{N}_{>0}$; $\mathbb{N}$ represents the set of all natural numbers, $\mathbb{N}_{>0}$ is, therefore, the set of all non-zero natural numbers.
\end{definition}

\begin{assumption}
Frequencies with lower order in the linear combinations are more likely to be observed in the evoked SSVEP.
\end{assumption}

With these two assumptions, we formulate \textit{Multi-Frequency CCA (MFCCA)}, which is 
explicitly designed to accommodate multi-frequency stimulation in SSVEP that satisfies these assumptions. Assumption 2 also allows the parameterisation of the algorithm in a similar way as CCA but, instead of bounding the reference set with number of harmonics $N_h$, we bound it with the order number $N_\mathrm{O}$.

\subsubsection{MFCCA}\label{sec:MFCCA}
Using Assumptions 1 and 2, the reference set can be adjusted to improve the robustness of CCA in handling multi-frequency stimulation.
The adjusted reference set consists of the sines and cosines of the positive frequencies that are integer linear combinations of the stimulation frequencies up to order $N_\mathrm{O}$. For example, for $N=2$ and $N_\mathrm{O}=2$, the reference set is
\begin{equation}\label{eq:y_MFCCA}
    y_{\textrm{MFCCA}\_N_\mathrm{O}=2}(t)=
    \begin{bmatrix}
        \sin(2\pi f_1 t)\\
        \cos(2\pi f_1 t)\\
        \sin(2\pi f_2 t)\\
        \cos(2\pi f_2 t)\\
        \sin(2\pi (2f_1) t)\\
        \cos(2\pi (2f_1) t)\\
        \sin(2\pi (2f_2) t)\\
        \cos(2\pi (2f_2) t)\\
        \sin(2\pi (f_1+f_2)t)\\
        \cos(2\pi (f_1+f_2)t)\\
        \sin(2\pi |f_1-f_2|t)\\
        \cos(2\pi |f_1-f_2|t)
    \end{bmatrix}.
\end{equation}

In case of $N_\mathrm{O}=3$, the sine and cosine terms with frequencies $3f_1$, $3f_2$, $2f_1+f_2$, $|2f_1-f_2|$, $f_1+2f_2$, and $|f_1-2f_2|$ will be appended to $y_{\textrm{MFCCA}\_N_\mathrm{O}=2}(t)$, assuming these frequencies are all positive. Similarly, the reference set can be determined and bounded with any given order $N_\mathrm{O}$.

The difference between the proposed and conventional techniques is in the construction of the reference set, where order is utilised to determined the terms in the reference sets. In conventional CCA, no interactions between the fundamental frequencies and harmonics are considered. Therefore, in the experiment, comparison is made between the proposed MFCCA and standard CCA with $N_h=N_\mathrm{O}$.

\subsection{Experiment}\label{sec:Experiment}

\subsubsection{Experimental setup}
As shown in Fig. \ref{fig:Experiment}, subjects were centred and positioned 100 cm away from the stimulus with a chin rest supporting their head to maintain a consistent distance throughout the experiment. The stimulus was a Adafruit NeoPixel 8 $\times$ 8 RGB LED panel (70 mm $\times$ 70 mm) mounted on medium-density fibreboard. The colour of the LEDs was set to red.

\begin{figure}
    \centering
    \vspace{0.2cm}
    \includegraphics[trim={0.2cm 5cm 2.5cm 1.5cm}, clip, width=0.9\linewidth, keepaspectratio]{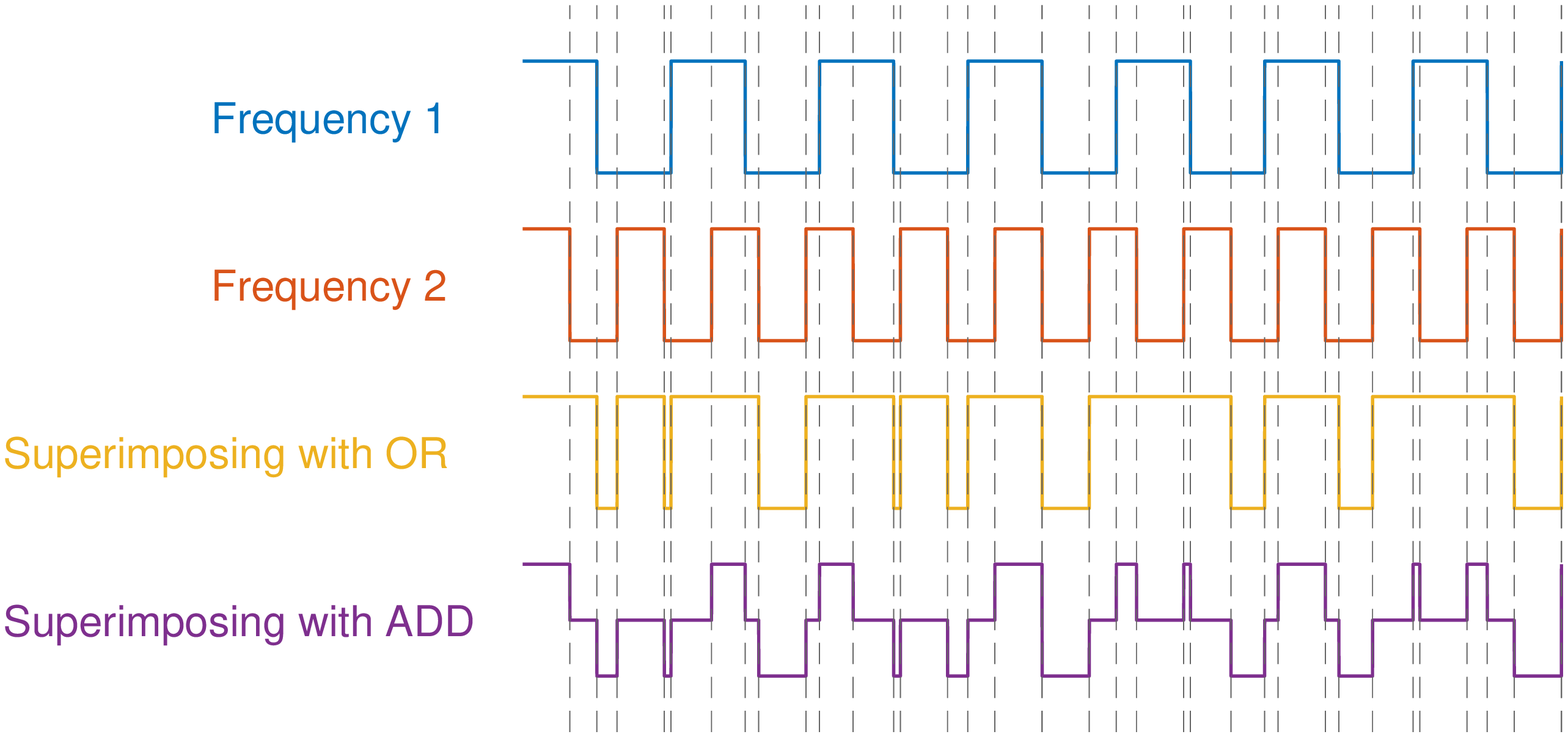}
    \caption{Frequency Superposition method with OR logic and two 50\% duty cycle square waves.}\vspace{-0.3cm}
    \label{fig:freqSup}
\end{figure}

Frequency superposition \cite{mu2021frequency} with OR logic and two 50\% duty cycle square waves, as shown in Fig. \ref{fig:freqSup}, was used as the stimulation method in the experiment. Frequency superposition method superimposes multiple independent input signals on the stimulation medium. With the OR logic and 50\% duty cycle square waves, stimulus will be set to full brightness if one of the inputs is HIGH.

\subsubsection{Experimental protocol}
Six frequency pairs were tested during the experiment: 7 \& 9~Hz, 7 \& 11~Hz, 7 \& 13~Hz, 9 \& 11~Hz, 9 \& 13~Hz, and 11 \& 13~Hz. This made a total of six trials with stimulation frequencies in the above mentioned sequence. With each frequency pair, a 10~s visual cue was first displayed, followed by a 30~s visual stimulation. A 20~s rest was provided after each trial.

\subsubsection{Data acquisition}
The g.USBamp amplifier with g.SAHARA dry electrodes (g.tec medical engineering GmbH, Austria) were used to capture SSVEP in this experiment. Six channels over the visual cortex (PO3, POz, PO4, O1, Oz, O2, international 10-10 system) were selected for EEG recording. Data was recorded at 512 Hz with 50 Hz notch filter and 0.5-60 Hz bandpass filter applied in the g.USBamp settings. Reference and ground electrodes were placed on the left and right mastoids, respectively.

\subsubsection{Subjects}
Nine healthy subjects (four females and five males, aged 22-33 years, mean 26.8 years, standard deviation 3.7) participated in the experiment. All subjects were right-handed and had normal or corrected-to-normal (with glasses) vision. Five subjects were na\"ive to SSVEP-based BCIs. 

This experiment was approved by the University of Melbourne Human Research Ethics Committee (Ethics ID 1851283). Written consent was received from all subjects prior to the experiments.

\subsubsection{Data processing}\label{sec:dataProcessing}

To extract patterns in multi-frequency stimulated SSVEP, the full 30~s of data from each trial was used. The data was first averaged across all channels and then subtracted from the measurement from Oz. This was then bandpass filtered between 1.8-60 Hz in \textsc{Matlab} (MathWorks, R2020a) with command \textit{bandpass} (type of impulse response was selected by \textsc{Matlab}; steepness: 0.85; stopband attenuation: 60). Lastly, fast Fourier transform (FFT) was calculated to show the frequency domain patterns.

The 30~s raw data were decoded with both CCA (Eq.~\eqref{eq:y_multi}) and MFCCA under different selections of order $N_\mathrm{O}$ as a performance comparison.

\section{Results}
\begin{figure}
    \centering
    \vspace{0.2cm}
    \includegraphics[trim={0cm 0cm 0cm 0cm}, clip, width=0.72\linewidth, keepaspectratio]{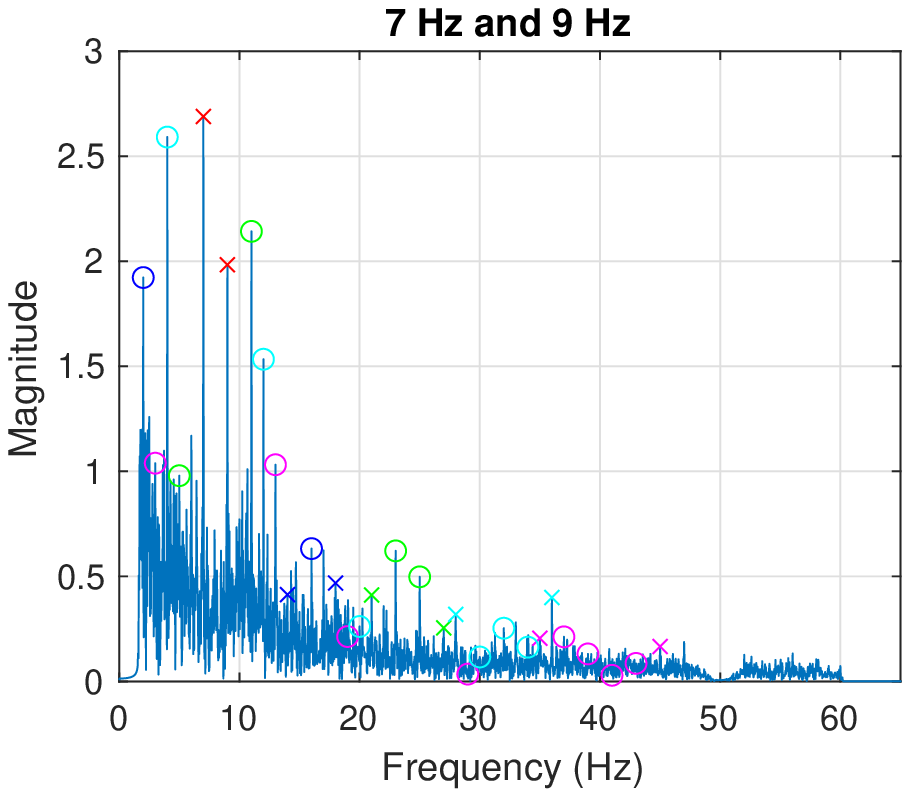}
    \caption{FFT magnitude plot of SSVEP recordings from subject 2 in the trial with 7 and 9 Hz stimulation. Red crosses label the two stimulation frequencies. Crosses and circles indicate harmonics and linear combinations of the two frequencies, respectively. Orders $N_\mathrm{O}$ are shown with different colours: red -- $N_\mathrm{O}=1$; blue -- $N_\mathrm{O}=2$; green -- $N_\mathrm{O}=3$; cyan -- $N_\mathrm{O}=4$; magenta -- $N_\mathrm{O}=5$.}\vspace{-0.3cm}
    \label{fig:Result}
\end{figure}

\begin{figure}
    \centering
    \includegraphics[trim={0cm 0cm 0cm 0cm}, clip, width=0.7\linewidth, keepaspectratio]{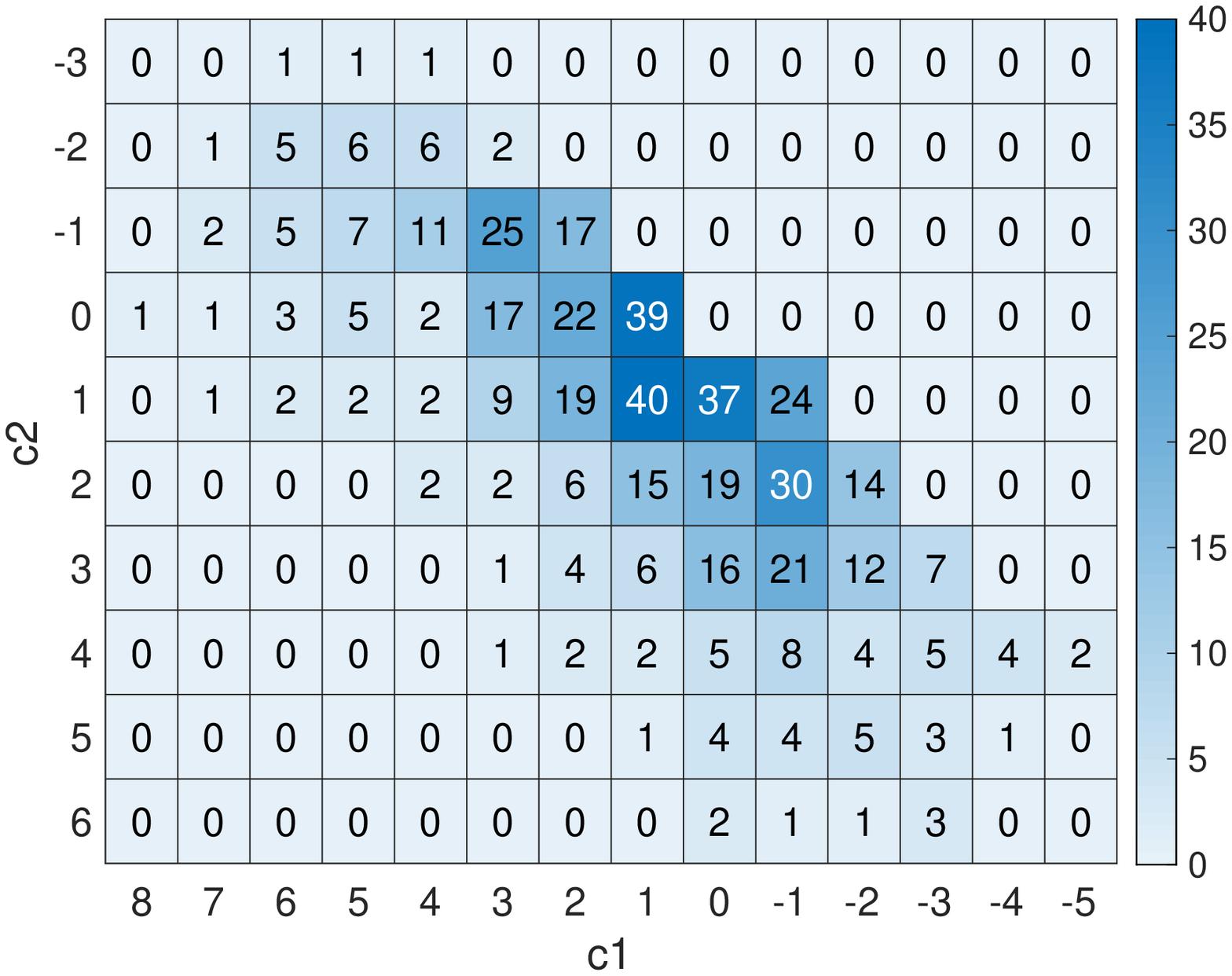}
    \label{fig:order_heatmap}
    \caption{Number of occurrences of each coefficient pair; $c1$ and $c2$ are the coefficients of the lower and higher frequencies, respectively.}\vspace{-0.3cm}
    \label{fig:order}
\end{figure}

The resulting SSVEP contain not only the input frequencies and their harmonics, but also the integer linear combinations of the two frequencies. Further details can be found in \cite{mu2021frequency}.
For example, Fig.~\ref{fig:Result} shows the FFT magnitude plot of the recorded SSVEP from Subject 2 in 7 and 9 Hz trial. The peaks have been labelled in different colours based on the order $N_\mathrm{O}$ defined in Section~\ref{sec:assumptions}.
Fig.~\ref{fig:order} shows the number of occurrences of each coefficient pair in the top ten peaks in each trial. It can be observed that the lower the order, the more likely it appeared in the peaks.
Assumptions 1 and 2 are validated by the results shown in Fig.~\ref{fig:Result} and ~\ref{fig:order}.

\begin{figure*}
    \centering
    \vspace{0.24cm}
    \includegraphics[trim={0cm 0cm 0cm 0cm}, clip, width=0.97\linewidth, keepaspectratio]{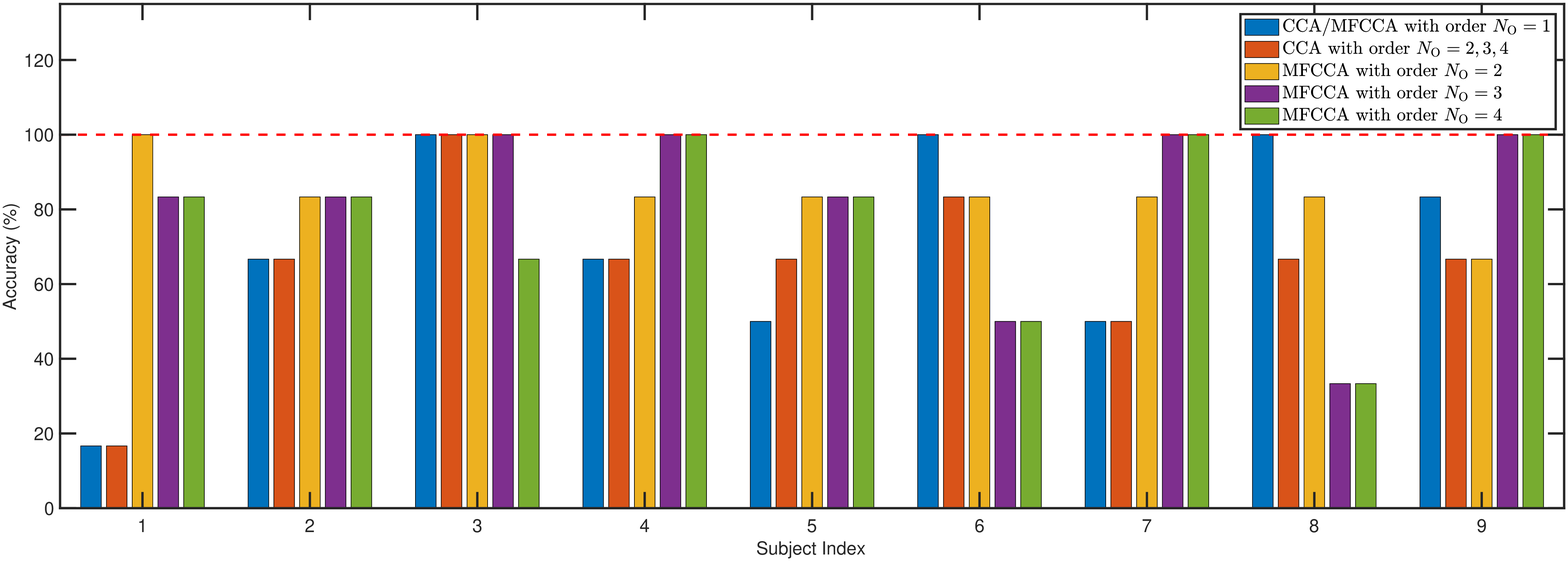}
    \caption{Accuracies for standard CCA and MFCCA for each subject. Results from CCA with orders 2, 3, and 4 are shown by the single blue bar for each subject, as there is no variation in accuracy under these settings. The orange, yellow, and purple bars show accuracies for MFCCA with orders 2, 3, and 4, respectively.}\vspace{-0.3cm}
    \label{fig:CCAvsMFCCA_perSubj}
\end{figure*}

\begin{figure}
    \centering
    \includegraphics[trim={0cm 0cm 0cm 0cm}, clip, width=0.75\linewidth, keepaspectratio]{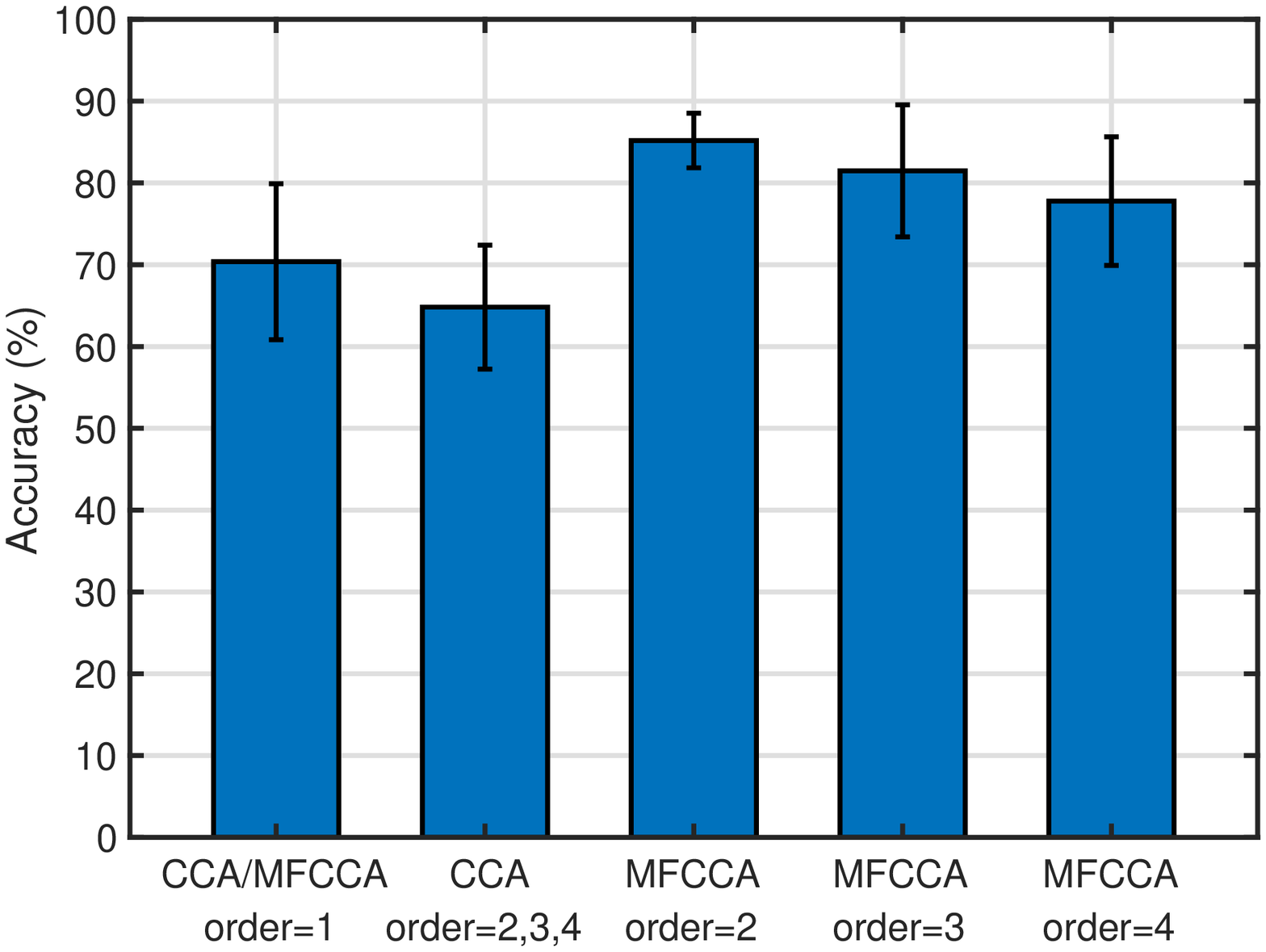}
    \caption{Comparisons between standard CCA and MFCCA. The heights of the bars are the mean accuracies in each decoder setting. Error bars show $\pm 1$ standard error. (CCA: standard CCA; MFCCA: multi-frequency CCA.)}\vspace{-0.3cm}
    \label{fig:CCAvsMFCCA_mean}
\end{figure}

Figs.~\ref{fig:CCAvsMFCCA_perSubj} and \ref{fig:CCAvsMFCCA_mean} show comparisons of decoding accuracy between CCA and MFCCA. In Fig.~\ref{fig:CCAvsMFCCA_perSubj}, it is observed that the highest accuracy usually occurred with MFCCA at orders 2 or 3. Different subjects showed different trends when the order was increased. However, for all subjects, MFCCA was always equal to or better than standard CCA when the order was set to 2.

The accuracies for each subject were grouped based on the decoder setting.  Comparisons of the mean accuracy are shown in Fig.~\ref{fig:CCAvsMFCCA_mean}. MFCCA produced higher classification accuracy compared to standard CCA at orders 2, 3, and 4. When the order was 1, the two formulations were the same and do not perform better than MFCCA at higher orders.

Based on the Jarque-Bera test, accuracies were normally distributed in all the five groups at $\alpha=0.05$ significance level. When order $N_\mathrm{O}=2$, CCA and MFCCA were significantly different based on the paired-sample t-test ($p=0.0471$). Statistical significance was not observed at $\alpha=0.05$ when $N_\mathrm{O}=3$ ($p=0.1838$) and $N_\mathrm{O}=4$ ($p=0.3360$).

\section{Discussion}

The experimental results validated the two assumptions underlying MFCCA and, therefore, justified the use of MFCCA in decoding multi-frequency stimulated SSVEP.

The decoding accuracy was higher when we included linear combinations of frequencies in the reference set, as shown in Fig. \ref{fig:CCAvsMFCCA_mean}.
A significant improvement in decoding accuracy could be seen with MFCCA compared with CCA when the order was 2. When the order was 3, six out of nine subjects showed improvement in decoding accuracy with MFCCA, and one subject remained at 100\% accuracy. The same six subjects also showed improvement in decoding accuracy when the order was 4.

We observed that, for two subjects (6 and 8), MFCCA did not perform as well at higher order. We believe this could be due to differences in SSVEP response in different subjects. In particular, these were two of the five subjects who were na\"ive to SSVEP-based BCIs. Since we did not include specific selection criteria in recruiting subjects, several outliers are considered acceptable.

Even though the performance of MFCCA varies with the selection of order and did not always perform as well at higher orders, MFCCA makes ``training-free'' BCIs possible using SSVEP with multi-frequency stimulation. On average, MFCCA improved decoding accuracy by 20\%, 17\%, and 13\% at orders 2, 3, and 4, respectively, compared to CCA.


\section{Conclusion}

In this paper, a novel extension of CCA, termed the Multi-Frequency CCA (MFCCA), is proposed that is explicitly designed to decode SSVEP response with multi-frequency stimulation. The proposed formulation is based on two assumptions that were experimentally validated in this paper. The concept of order was introduced and used as a tuning parameter in the decoding algorithm.
With the interactions between the multiple stimulation signals of different frequencies taken into consideration and bounded by order, the performance of MFCCA was observed to vary with the choice of order with decreased performance observed on higher orders. The peak performance was observed in this experiment at order 2 resulting in a 20\% better performance when compared to standard CCA.



\bibliographystyle{IEEEtran}
\bibliography{IEEEabrv,ref}

\end{document}